\documentclass[12pt]{article}
\usepackage{makeidx}
\usepackage{graphics}
\usepackage{amsthm}
\usepackage{amssymb}
\usepackage{amsmath}
\usepackage{amsfonts}
\usepackage{epic}
\usepackage{hyperref}
\usepackage{epsfig}
\usepackage{bm}
\usepackage{amscd}

\theoremstyle{plain}

\theoremstyle{definition}

\def\be{\begin{equation}}
\def\ee{\end{equation}}
\smallskip
\input{tcilatex}

\begin{document}


\begin{titlepage}
\begin{flushright}
\end{flushright}
\begin{center}
\noindent{{\LARGE{A Note on ${\mathbb{Z}}_2$ Symmetries of the KZ Equation}}} 

\smallskip

\smallskip


\smallskip

\smallskip
\smallskip
\noindent{\large{Gaston E. Giribet}}
\end{center}
\smallskip

\smallskip
\smallskip

\smallskip

\centerline{Departamento de F\'{\i}sica, Universidad de Buenos Aires, FCEN UBA}
\centerline{{\it Ciudad Universitaria, Pabell\'on I, 1428, Buenos Aires, Argentina}}
\smallskip

\smallskip

\smallskip
\smallskip
\smallskip

\begin{abstract}
We continue the study of hidden ${\mathbb{Z}}_2$ symmetries of the
four-point $\hat {sl(2)} _k$ Knizhnik-Zamolodchikov equation iniciated in \cite{Yo2005}. Here, we focus our attention on
the four-point correlation function in those cases where one spectral flowed state of
the sector $\omega =1$ is involved. We give a formula that shows how this observable 
can be expressed in terms of the four-point
function of non spectral flowed states. This means that the formula holding for the winding violating 
four-string scattering processes in $AdS_3$ has a simple expression in terms of the one for the 
conservative case, generalizing what is
known for the case of three-point functions, where the violating and the
non-violating structure constants turn out to be connected one to each other in a 
similar way. What makes this connection particularly simple is the fact that, unlike 
what one would naively expect, it is not necessary to explicitly solve the five-point function 
containing a single spectral flow operator to this end. Instead, non diagonal functional relations between 
different solutions of the KZ equation turn out to be the key point for this short path to exist. Considering 
such functional relation is necessary but it is not sufficient; besides, the formula also follows from the relation 
existing between correlators in both WZNW and Liouville conformal theories.

\end{abstract}

\end{titlepage}



\newpage


\section{Introduction}

The $SL(2,\mathbb{R})_{k}$ WZNW model plays an important role within the
context of string theory. This conformal model enters in the worldsheet
description of the theory formulated on exact non-compact curved
backgrounds, being the most celebrated examples: the string theory on $%
AdS_{3}$ and on the 2D black hole (the last, by means of its gauged $SL(2,%
\mathbb{R})_{k}/U(1)$ construction). Consequently, among the main
motivations, this topic received particular attention due to its relevance
for black hole physics and its relation with the $AdS/CFT$ correspondence.
Actually, this conformal field theory raises the hope to work out the
details of the correspondence beyond the particle limit approximation \cite%
{GKS,GK,Oetal}. Here, we continue the study of the observables in this
theory, focussing out attention on the four-point function.

\subsection{String amplitudes in $AdS_{3}$}

Although the structure of the WZNW model on $SL(2,\mathbb{R})$
(corresponding to strings in Lorentzian $AdS_{3}$) is not yet well
understood, the observables of this theory are assumed to be well defined in
terms of the analytic continuation of the correlation functions in the
Euclidean case, i.e. in the $SL(2,\mathbb{C})_{k}/SU(2)$ WZNW gauged model.
Hence, the string scattering amplitudes in Lorentzian $AdS_{3}$ are obtained
by integrating over the worldsheet insertions of vertex operators in the
model on $SL(2,\mathbb{C})/SU(2)$ and then extending the range of the
indices of continuous representations in order to include the discrete
representations of $SL(2,\mathbb{R})$ as well. The correlation functions
defined in such a way typically develop poles in the space of representation
indices and these poles are then interpreted as state conditions of bounded
states (called \textquotedblleft short strings" in \cite{MO3}) and yield
selection rules \cite{SHope,Sope} for the string scattering amplitudes.
Besides, the continuous representations (the \textquotedblleft long strings"
in \cite{MO3}) have a clear interpretation as asymptotic states and define
the S-matrix in the Lorentzian target space.

In the last decade, we gained important information on the explicit
functional form of the WZNW correlation functions, and this enabled us to
study the string theory on $AdS_{3}$ beyond the supergravity approximation,
both at the level of the spectrum \cite{MO1} and at the level of its
interactions (see \cite{GN3,MO3} and references therein). Originally, some
particular cases of the two and three-point functions were explicitly
computed in \cite{B,BB}, and was Teschner who presented the general result
in \cite{Tboost} and carefully described its formal aspects in \cite%
{Tboost,Tmini,Tope}, by employing the boostrap approach and the
minisuperspace approximation. Subsequently, other approaches, like the path
integral techniques \cite{Ponjaspath} and the free field representation \cite%
{GN2,Ponjasfree}, were employed to rederive such correlation functions. The
operator product expansion was studied in Ref. \cite{SHope,Sope} in order to
investigate the fusion rules, and the crossing symmetry of the correlation
functions was eventually proven in Ref. \cite{Tcross} by making use of its
analogy with the Liouville field theory \cite{FZ,Andreev}; see also \cite%
{Ponsot}. The study of the correlation functions was also shown to be useful
for consistency checks of the conjectured dualities between this and other
conformal models \cite{FH,GKnotes,KKK}.

After these objectives were achieved, the study of correlation functions in
the $SL(2,\mathbb{R})_{k}$ WZNW model acquired a new dimension since
Maldacena and Ooguri pointed out the existence of new representations of $%
SL(2,\mathbb{R})_{k}$ contributing to the spectrum of the string theory in $%
AdS_{3}$ \cite{MO3}. Then, the correlation functions involving these new
states had to be analyzed as well. These new representations, obtained from
the standard ones by acting with the spectral flow transformation, are
semiclassically related to the possibility of the $AdS_{3}$ strings to have
a nonzero winding number. This interpretation in terms of \textquotedblleft
winding numbers" does not regard a topological winding, but it turns out to
be a consequence of the presence of a non trivial NS-NS $B_{\mu \nu }$
background field. Then, from the beginning, this winding number, as a non
topological degree of freedom, was assumed to be possibly violated when the
interactions were to take place. This violation was actually first suggested
in a unpublished work by Fateev and the brothers Zamolodchikov for the case
of the 2D black hole \cite{FZZ}. A free field computation of three-point
function of such winding states, including the violating winding case,
appeared in Ref. \cite{GN3}, and a similar free field realization was
studied in more detail in \cite{GN2,GLopez}. An impressive analysis of the
correlation functions in the $SL(2,\mathbb{R})_{k}$ WZNW model was then
presented in the paper \cite{MO3}, by Maldacena and Ooguri. There, the pole
structure of two, three and four-point functions was discussed in the
framework of the semiclassical interpretation and the $AdS/CFT$
correspondence. The exact expressions of two and three-point functions,
including the violating winding three-point function, were fully analyzed.
Besides, the string theory interpretation of such observables, as
representing scattering amplitudes, was given with precision. The four-point
function was also studied in \cite{MO3}, and, even though there is still no
closed expression for the generic case available, our understanding of its
analytic structure was substantially increased due to the results of that
work. By making use of the factorization \textit{ansatz} given by Teschner
in \cite{Tope}, Maldacena and Ooguri proposed an analytic extension of the
expression for the $SL(2,\mathbb{C})_{k}/SU(2)$ conformal blocks. Thus, they
gave a precise prescription to integrate the monodromy invariant expression
over the space of $SL(2,\mathbb{R})_{k}$ representations and to pick up the
pole contribution of the discrete states. Perhaps, the two main observations
made in \cite{MO3} regarding the four-point functions are: the existence of
additional poles in the middle of the moduli space, and the fact that the
factorization of the four-point function only permits the usual
interpretation in terms of physical intermediate states for particular
incoming and outgoing kinematic configurations. The analytic structure of
particular four-point functions, those leading to logarithmic singularities,
was studied in \cite{GSimeone}, showing that this example fits the standard
structure of four-point functions in the $AdS/CFT$ correspondence.

Despite all this information we get about four-point functions, it is worth
noticing that the mentioned cases only took into account non-flowed
representations (representing non-winding string states) as those describing
the incoming and outgoing states. For instance, even though the winding
strings of the sector $\omega =-1$ were shown to arise in the intermediate
channels of four-point functions \cite{MO3}, this was shown by analyzing the
processes that only involve incoming and outgoing states of the sector $%
\omega =0$. Then, it would be interesting to extend the study to the case of
four-point functions that involve external states of the sectors $\omega
\neq 0$. Here is where our result enters in the game since it actually
permits to get information of the four-point winding violating functions
from all what is already known about the conservative case. In fact, in this
note we will show how to connect the correlation functions involving one
flowed state (winding string state) of the sector $\omega =-1$ (with winding
number $\omega =-1$) to the analogous quantity that merely involves
non-flowed states (just strings with winding number $\omega =0$). We will
also argue that this is actually analogous to the relation existing between
the violating winding three-point function and the conservative one. The way
of showing such connection takes into account a recent result that presents
a new map between WZNW and Liouville correlators. This map, different from
that employed in \cite{Tope} to prove the crossing symmetry in $SL(2,\mathbb{%
C})/SU(2)$, was discovered by Stoyanovsky some years ago \cite{S2000}, and
it was further developed by Teschner and Ribault in Ref. \cite%
{RT2005,R2005,R2005b}. In \cite{GNakayama2005,chinitos,chinitos2,chinitos3}
this map between both CFTs (henceforth denoted SRT map) was analyzed in the
context of the implications it has in string theory (see also Ref. \cite%
{Related} for recent works on the relation between Liouville and WZNW
correlators). In references \cite{Yo2005b,Yo2006}, a free field realization
of the SRT map was given, and it was shown to reproduce the correct
three-point function for the case where one spectral flowed state (string
state with winding $\omega =1$) is considered. Such observable had been also
computed in \cite{Yo2005} through \textquotedblleft the
other\textquotedblright\ connection to Liouville theory, discovered by
Fateev and Zamolodchikov in \cite{FZ} and extended in Ref. \cite%
{Andreev,Tope,Ponsot} to the non-compact case (henceforth denoted FZ map).
We will be more precise in the following paragraph.

\subsection{Overview}

Recently, a new possibility to study the four-point function involving
winding states has raised because of a discovery made by Ribault \cite{R2005}
and Fateev \cite{Funpublished}, stating that correlators involving winding
states in WZNW can be written in terms of correlators in Liouville field
theory. In the case of four-point function with one state in the sector $%
\omega =-1$, on which we are interested here, this turns out to correspond
to the Liouville five-point function. In principle, this does not seem to
imply an actual simplification since five-point function in Liouville theory
cannot be simply solved either. However, we noticed that the SRT map is not
the only way of mapping \textquotedblleft the same\textquotedblright\
Liouville five-point function to a (different) four-point function of the
WZNW theory. Indeed, we can also do this by employing the non-compact
generalization of the FZ map \cite{FZ}. If this \textquotedblleft second
map\textquotedblright\ is used, the four-point function reached in the WZNW
side is one that contains four non-winding states, enabling us to relate
winding violating processes in $AdS_{3}$ with their conservative analogs.
However, this is not the whole story since, unlike the SRT map \cite%
{S2000,RT2005,R2005}, the FZ map involves a non-diagonal correspondence
between the quantum numbers of both Liouville and WZNW sides. Hence, besides
the appropriate combination of the SRT and the FZ maps, a sort of a
\textquotedblleft diagonalization procedure" is also required in order to
present the result in a clear form. This diagonalization is eventually
achieved by using the FZ map itself and the reflection symmetry of Liouville
theory. By doing something similar to that in Ref. \cite{Yo2005}, we will
first indicate how such diagonalization is realized due to identities
holding between different exact solutions of the KZ equation. In fact, this
paper can be regarded as a addendum to Ref. \cite{Yo2005}, being the second
part of our study of hidden symmetries in the four-point $\widehat{sl(2)}%
_{k} $ KZ equation. In \cite{Yo2005}, a set of $\mathbb{Z}_{2}$ symmetry
transformations of the KZ\ equation were studied. Such involutions were
realized by means of the action on the four indices of $SL(2,\mathbb{R})$
representations and led to prove identities between different exact
solutions of the KZ equation. The main tool for working out such identities
was the FZ map, mapping four-point functions of the WZNW theory to a
particular subset of five-point functions of the Liouville field theory.
Following this line, here we explore the implications of new symmetry
transformations on the solutions of KZ\ equation. The plan of the paper goes
as it follows: In the next section we will review the connections between
four-point functions in WZNW theory and five-point function in Liouville
field theory. The fact that there is not a unique map of this kind\footnote{%
or, more precisely, the fact that the connection between the different maps
which are known turns out to be non trivial, see Ref. \cite{RT2005}.} is the
reasons for a non trivial relation between correlators of winding and
non-winding states to exist. In section 3, as a preliminary result, we first
prove a new identity between exact solutions of the KZ equation,
complementing the catalog presented in Ref. \cite{GSimeone, Yo2005}. This
identity is again realized by a non diagonal $\mathbb{Z}_{2}$ transformation
of the class studied in Ref. \cite{Yo2005}, acting on the four indices of
the representations of $SL(2,\mathbb{R})$. Then, this leads us to show how
the four-point correlation function involving one spectral flowed state in
the winding sector $\omega =-1$ can be written in terms of the correlation
function of non flowed (non winding) states.

\section{Conformal field theory}

\subsection{The WZNW model}

We are interested in the WZNW model formulated on the $SL(2,\mathbb{R})$
group manifold. Its action corresponds to the non-linear $\sigma $-model of
strings in Lorentzian $AdS_{3}$ space. On the other hand, its Euclidean
version is similarly given by the gauged $SL(2,\mathbb{C})/SU(2)$ model. The
states of the Euclidean model are characterized by normalizable operators on
the Poincar\'{e} hyper half-plane $H_{3}^{+}=SL(2,\mathbb{C})/SU(2)$. These
operators can be conveniently written as%
\begin{equation}
\Phi _{j}(x|z)=\frac{1-2j}{\pi }\left( u^{-1}+u|\gamma -x|^{2}\right) ^{-2j},
\label{OPahora}
\end{equation}%
where the variables $\gamma ,\overline{\gamma }$ and $u$ are associated to
the $AdS_{3}$ Euclidean metric in Poincar\'{e} coordinates; namely%
\begin{equation*}
ds^{2}=k(u^{-2}du^{2}+u^{2}d\gamma d\overline{\gamma });
\end{equation*}%
while the complex coordinates $x$ and $\overline{x}$ represent auxiliary
variables that expand the $SL(2,\mathbb{C})$ representations as it follows%
\begin{equation*}
J^{a}(z_{1})\Phi _{j}(x|z_{2})\sim \frac{1}{z_{1}-z_{2}}D_{x}^{a}\Phi
_{j}(x|z_{2})+...
\end{equation*}%
for $a=\{3,+,-\}.$ The dots \textquotedblleft ..." stand for
\textquotedblleft regular terms" in the OPE, while the differential
operators $D_{x}^{a}$ correspond to the realization 
\begin{equation*}
D_{x}^{3}=x\partial _{x}+j,~\qquad D_{x}^{+}=\partial _{x},~\qquad
D_{x}^{-}=x^{2}\partial _{x}+2jx.
\end{equation*}%
Besides, the operators $J^{a}(z)$ are the local Kac-Moody currents, whose
Fourier modes are defined by $J^{a}(z)=\sum_{n\in \mathbb{Z}}J_{n}^{a}$ $%
z^{-1-n}$ and satisfy the $\widehat{sl(2)}_{k}$ affine Kac-Moody algebra of
level $k$. The value of $k$ is related to the string length and the $AdS$
radius through $k=l_{AdS}^{2}/l_{s}^{2}$. This keeps track of the conformal
invariance of the WZNW theory. The Sugawara construction yields the
stress-tensor that can be used to compute the central charge of this theory,
being 
\begin{equation}
c=3+\frac{6}{k}.
\end{equation}%
The formula for the conformal dimension of operators (\ref{OPahora}) reads%
\begin{equation}
h_{j}=\frac{j(1-j)}{k-2},  \label{dddd}
\end{equation}%
where the indices take the values $j=\frac{1}{2}+i\mathbb{R}_{>0}$ for $SL(2,%
\mathbb{C})/SU(2)$. In order to propose a similar algebraic realization for
the Lorentzian string theory, it is usually assumed that the observables of
the Euclidean theory admits an analytic continuation in the variable $j$,
now parameterizing both continuous and discrete representations of $SL(2,%
\mathbb{R})$. It is worth noticing that the transformation $j\rightarrow 1-j$
is a symmetry of the formula (\ref{dddd}); this corresponds to the so called
Weyl reflection symmetry and will be important for us in section 3.
Operators (\ref{OPahora}) represent the vertex operators of the string
theory in $AdS_{3}$ and define the correlation functions. Then, after
integrating over the variables $z_{\mu }$, the functional form of the $N$%
-point correlators will still depend on the auxiliary variables $x_{\mu }$,
with $\mu =\{1,2,...N\}$. Within the context of the $AdS_{3}/CFT_{2}$
interpretation, those auxiliary variables are interpreted as the coordinates
of the dual conformal field theory in the boundary of $AdS_{3}$, and then
acquire a geometrical meaning. This picture, employing the variable $x$ to
label the representations, is usually referred as the $x$-basis. On the
other hand, there exists a different picture that is also convenient, and in
which the quantum numbers labeling the representations permits to define
string states with well defined momenta in the bulk. This is the often
called $m$-basis, and employs the standard way of parameterizing
representations of the $SL(2,\mathbb{R})$ group, by using a pair of indices $%
j,m$. In this frame, the string states in $AdS_{3}$ are given by vectors $%
|j,m>\otimes $ $|j,\overline{m}>$ of the $SL(2,\mathbb{R})\times SL(2,%
\mathbb{R})$ representations. Furthermore, as it was mentioned in the
introduction, in Ref. \cite{MO3} it was shown that an additional quantum
number should be included in order to fully characterize the space of states
in $AdS_{3}$. This quantum number, denoted $\omega $, is associate to the
winding number of strings in $AdS_{3}$, at least in what respects to the
states with a suitable asymptotic description (the long strings). From an
algebraic point of view, $\omega $ labels the spectral flow transformation
that generates new representations of the theory. This is the reason because
we will refer to the \textquotedblleft spectral flowed states" as
\textquotedblleft winding states", indistinctly. The vertex operators
representing winding states (states with $\omega \neq 0$) in the $m$-basis
are to be denoted $\Phi _{j,m,\overline{m}}^{\omega }(z)$ and define the
correlation functions in this basis, which we will denote as $\left\langle
\Phi _{j_{1},m_{1},\overline{m}_{1}}(z_{1})\Phi _{j_{2},m_{2},\overline{m}%
_{2}}(z_{2})...\Phi _{j_{N},m_{N},\overline{m}_{N}}(z_{N})\right\rangle .$
The conformal dimension of the states in the $m$-basis depends both on $m$
and $\omega $ through the expression%
\begin{equation*}
h_{j,m,\omega }=\frac{j(1-j)}{k-2}-m\omega -\frac{k}{4}\omega ^{2}\text{.}
\end{equation*}%
In the case $\omega =0,$ operators $\Phi _{j,m,\overline{m}}^{\omega =0}(z)$
are related to those of the $x$-basis through the Fourier transform 
\begin{equation}
\Phi _{j,m,\overline{m}}^{\omega =0}(z)=\int d^{2}x\ \Phi _{j}(x|z)x^{m-j}%
\overline{x}^{\overline{m}-j}  \label{Fourier}
\end{equation}%
On the other hand, the definition of string states with $\omega \neq 0$ in
the $x$-basis was studied in \cite{MO3,Minces}; however, these have not a
simple expression. Now, let us discuss the correlation functions in more
detail.

\subsection{The four-point KZ equation}

As it was commented, the two and three-point functions in the WZNW model are
known \cite{BB}; and the four-point function in the sector $\omega =0$ was
studied in detail in Ref. \cite{Tope}, where a consistent $\mathit{ansatz}$
was proposed based in the analogy with other CFTs; we detail such proposal
below.

The four-point correlation functions on the zero-genus topology are
determined by conformal invariance up to a factor $f$, which is a function
of the cross ratio $z$ and the variables $j_{i},x_{i}$ and $\bar{x}_{i},$
that label the representations; namely%
\begin{equation}
\left\langle \Phi _{j_{1}}(x_{1}|z_{1})\Phi _{j_{2}}(x_{2}|z_{2})\Phi
_{j_{3}}(x_{3}|z_{3})\Phi _{j_{4}}(x_{4}|z_{4})\right\rangle
=\prod_{a<b}^{4}|x_{a}-x_{b}|^{2J_{ab}}%
\prod_{a<b}^{4}|z_{a}-z_{b}|^{2h_{ab}}|f_{j_{1},j_{2},j_{3},j_{4}}(x,z)|^{2}
\label{recorda}
\end{equation}%
being 
\begin{equation*}
x=\frac{(x_{2}-x_{1})(x_{3}-x_{4})}{(x_{4}-x_{1})(x_{3}-x_{2})},\qquad z=%
\frac{(z_{2}-z_{1})(z_{3}-z_{4})}{(z_{4}-z_{1})(z_{3}-z_{2})},
\end{equation*}%
and where $h_{34}=h_{1}+h_{2}-h_{3}-h_{4},$ $h_{14}=-2h_{1},$ $%
h_{24}=h_{1}-h_{2}+h_{3}-h_{4},$ $h_{23}=h_{4}-h_{1}-h_{2}-h_{3};$ and $%
J_{34}=j_{1}+j_{2}+j_{3}-j_{4},$ $J_{14}=-2j_{1},$ $%
J_{24}=j_{1}-j_{2}+j_{3}-j_{4},$ $J_{13}=-j_{1}-j_{2}-j_{3}+j_{4}$.

The function $f=f_{j_{1},j_{2},j_{3},j_{4}}(x,z)$ is then given by certain
linear combination of solutions to the Knizhnik-Zamolodchikov partial
differential equation (KZ); \textit{i.e.} that combination which turns out
to be monodromy invariant. The KZ equation in the case of the $SL(2,\mathbb{R%
})_{k}$ WZNW model takes the form 
\begin{equation}
(k-2)z(z-1)\frac{\partial }{\partial z}f_{j_{1},j_{2},j_{3},j_{4}}(x,z)=%
\left( (z-1){\mathcal{D}}_{1}+z{\mathcal{D}}_{0}\right)
f_{j_{1},j_{2},j_{3},j_{4}}(x,z)  \label{balinf}
\end{equation}%
where the differential operators are 
\begin{eqnarray*}
{\mathcal{D}}_{1} &=&x^{2}(x-1)\frac{\partial ^{2}}{\partial x^{2}}-\left(
(j_{4}-j_{3}-j_{2}-j_{1}-1)x^{2}+2j_{2}x+2j_{1}x(1-x)\right) \frac{\partial 
}{\partial x}+ \\
&&+2(j_{1}+j_{2}+j_{3}-j_{4})j_{1}x-2j_{1}j_{2} \\
{\mathcal{D}}_{0} &=&-(1-x)^{2}x\frac{\partial ^{2}}{\partial x^{2}}+\left(
(-j_{1}-j_{2}-j_{3}+j_{4}+1)(1-x)-2j_{3}-2j_{1}x\right) (x-1)\frac{\partial 
}{\partial x}+ \\
&&+2(j_{1}+j_{2}+j_{3}-j_{4})j_{1}(1-x)-2j_{1}j_{3}
\end{eqnarray*}%
With \cite{Tope}, we can consider the following \textit{ansatz} for the
solution 
\begin{equation}
f_{j_{1},j_{2},j_{3},j_{4}}(x,z)=\int_{{\mathcal{C}}}dj\frac{%
C(j_{1},j_{2},j)C(j,j_{3},j_{4})}{B(j)}{\mathcal{G}}%
_{j_{1},j_{2,}j,j_{3},j_{4}}(x|z)\times \bar{{\mathcal{G}}}%
_{j_{1},j_{2,}j,j_{3},j_{4}}(\bar{x}|\bar{z}),  \label{antonia}
\end{equation}%
where the functions $C(j_{1},j_{2},j_{3})$ and $B(j_{1})$ are given by the
structure constants and the reflection coefficient of the $SL(2,\mathbb{R}%
)_{k}$ WZNW model, respectively; and where the contour of integration is
defined as covering the curve ${\mathcal{C}}=\frac{1}{2}+i\mathbb{R}$. The
integration along ${\mathcal{C}}$ turns out to be redundant for a monodromy
invariant solution since such particular linear combination is invariant
under Weyl reflection $j\rightarrow 1-j$, for which the contour transforms
as the complex conjugation ${\mathcal{C}}\rightarrow \overline{{\mathcal{C}}}%
=-{\mathcal{C}}$. Here, we are interested in the relation between different
solutions to (\ref{balinf}).

\subsection{The Liouville field theory}

Now, let us briefly review the Liouville field theory, which is the other
CFT in which we are interested here. Its action reads

\begin{equation}
S=\frac{1}{4\pi }\int d^{2}z\left( -\partial \varphi \bar{\partial}\varphi
+QR\varphi +\mu e^{\sqrt{2}b\varphi }\right) ,  \label{action}
\end{equation}%
where the background charge is given by $Q=b+b^{-1}$, and the exponential
self-interaction thus corresponds to a marginal deformation. We will set the
Liouville cosmological constant $\mu $ to have an appropriate value in order
to make the correlation functions acquire a simple form. This is achieved by
properly rescaling the zero mode of $\varphi (z)$. The Liouville field
theory is reviewed with impressive detail in Ref. \cite{Yu}; see also \cite%
{Tliouville,Seiberg} and the recent \cite{Leo}. The central charge of the
theory is given by%
\begin{equation*}
c=1+6Q^{2}>1
\end{equation*}%
and the vertex operators have the exponential form \cite{Tvertex}%
\begin{equation}
V_{\alpha }(z)=e^{\sqrt{2}\alpha \phi (z)},  \label{cinc}
\end{equation}%
whose conformal dimension is given by%
\begin{equation}
\Delta _{\alpha }=\alpha (Q-\alpha ).  \label{cinci}
\end{equation}%
Vertex operators (\ref{cinc}) define the Liouville $N$-point correlation
functions, which are to be denoted by $\left\langle
V_{a_{1}}(z_{1})V_{a_{2}}(z_{2})...V_{a_{N}}(z_{N})\right\rangle $. Notice
that formula (\ref{cinci}) remains invariant under $\alpha \rightarrow
Q-\alpha $, henceforth called Liouville reflection. This symmetry induces
the identification between both fields $V_{\alpha }(z)$ and $V_{Q-\alpha
}(z) $, yielding the operator valued relation%
\begin{equation}
\left\langle
V_{a_{1}}(z_{1})V_{a_{2}}(z_{2})...V_{a_{N}}(z_{N})\right\rangle
=R_{b}(\alpha _{1})\left\langle
V_{Q-a_{1}}(z_{1})V_{a_{2}}(z_{2})...V_{a_{N}}(z_{N})\right\rangle ,
\label{whatt}
\end{equation}%
which is valid for any vertex $i=\{1,2,...N\}$ (though we exemplified it
here for $i=1$), and where $R_{b}(\alpha _{1})$ represents the Liouville
reflection coefficient%
\begin{equation}
R_{b}(\alpha )=\left( \pi \mu \frac{\Gamma (1-b^{2})}{\Gamma (1+b^{2})}%
\right) ^{\frac{2}{b}\alpha -1-b^{-2}}\frac{\Gamma (2b\alpha -b^{2})\Gamma
(2b^{-1}\alpha -b^{-2})}{\Gamma (2-2b\alpha +b^{2})\Gamma (2-2b^{-1}\alpha
+b^{-2})},  \label{Reflection}
\end{equation}

Another important feature of the Liouville correlation functions is the fact
that those involving states with momentum $\alpha =-1/2b$ satisfy a well
known partial differential equation, called the
Belavin-Polyakov-Zamolodchikov equation (BPZ). This is similar to that of
the minimal models and actually holds for a wider family of correlators,
namely all of those involving certain state with momentum $\alpha _{m,n}=%
\frac{1-m}{2}b+\frac{1-n}{2}b^{-1}$ for any pair $m,n\in \mathbb{Z}_{>0}$.
In particular, here we are interested in Liouville five-point correlators of
the form%
\begin{equation*}
\left\langle
V_{a_{1}}(z_{1})V_{a_{2}}(z_{2})V_{a_{3}}(z_{3})V_{a_{4}}(z_{4})V_{a_{1,2}=-1/2b}(z_{5})\right\rangle .
\end{equation*}

Now, once both WZNW and Liouville theories were introduced, we move to the
following ingredient in our discussion: the close relation existing between
correlation functions of each of these two conformal theories. Such relation
has multiple aspects indeed; we will discuss two of them in the following
two subsections.

\subsection{The Fateev-Zamolodchikov identity}

The often called FZ map is a dictionary that connects four-point correlation
functions in WZNW models to five-point correlation in Liouville field
theory. This result was developed in Ref. \cite{Tcross} (see also Ref. \cite%
{Ponsot}) and, among other ingredients involved in its derivation, is based
on the relation existing between solution of the KZ and the BPZ differential
ecuations. Such relation was originally noticed by Fateev and Zamolodchikov
in Ref. \cite{FZ} for the case of the compact $SU(2)_{k}$ WZNW case and the
minimal models, and it basically states that, starting from any given
solution of the four-point KZ equation, a systematical way of constructing a
solution of the five-point BPZ equation exists. The FZ map, or strictly
speaking its adaptation to the non-compact WZNW model, was employed to
investigate several properties of the $SL(2,\mathbb{R})_{k}$ conformal
theory, becoming one of the most fruitful tools to this end. In particular,
Teschner gave to it its closed form and used it to prove the crossing
symmetry of the WZNW model by assuming that a similar relation holds for the
conformal blocks. Besides, Andreev rederived the fusion rules of admissible
representations of the $\hat{sl(2)}_{k}$ affine algebra by similar means 
\cite{Andreev}, and Ponsot discussed the monodromy of the theory with such
techniques \cite{Ponsot}. In Ref. \cite{Yo2005}, the FZ map was shown to be
useful to prove several identities between exact solutions of the KZ
equation, and we will extend such result in the next section here. But
first, let us briefly review the FZ statement: The observation made in \cite%
{FZ} is that the KZ equation satisfied by the four-point functions in the
WZNW model agrees with the BPZ equation satisfied by a particular set of
five-point functions in Liouville field theory. More specifically, Fateev
and Zamolodchikov showed that it is possible to get a solution of the KZ
equation by starting with one of the BPZ\ system. This yields the relation

\begin{equation*}
\langle \Phi _{j_{2}}(0|0)\Phi _{j_{1}}(x|z)\Phi _{j_{3}}(1|1)\Phi
_{j_{4}}(\infty |\infty )\rangle
=F_{k}(j_{1},j_{2},j_{3},j_{4})X_{k}(j_{1},j_{2},j_{3},j_{4}|x,z)\times
\end{equation*}%
\begin{equation}
\times \langle V_{\alpha _{2}}(0)V_{\alpha _{1}}(z)V_{\alpha _{3}}(1)V_{-%
\frac{1}{2b}}(x)V_{\alpha _{4}}(\infty )\rangle  \label{fia}
\end{equation}%
where%
\begin{equation*}
X_{k}(j_{1},j_{2},j_{3},j_{4}|x,z)=|z|^{-4b^{2}j_{1}j_{2}+\alpha _{1}\alpha
_{2}}|1-z|^{-4b^{2}j_{1}j_{3}+\alpha _{1}\alpha _{3}}|x-z|^{-2b^{-1}\alpha
_{1}}|x|^{-2b^{-1}\alpha _{2}}|1-x|^{-2b^{-1}\alpha _{3}},
\end{equation*}%
\begin{equation}
F_{k}(j_{1},j_{2},j_{3},j_{4})=c_{b}\left( \lambda \pi b^{2b^{2}}\frac{%
\Gamma (1-b^{2})}{\Gamma (1+b^{2})}\right)
^{1-3j_{1}-j_{2}-j_{3}-j_{4}}\left( \lambda \mu \right) ^{2j_{1}}\prod_{\mu
=1}^{4}\frac{\Upsilon _{b}(2j_{\mu }+1)}{\Upsilon _{b}(2\alpha _{\mu })},
\label{fia2}
\end{equation}%
and where the function $\Upsilon _{b}(x)$ is the one introduced in Ref. \cite%
{ZZ} when presenting the explicit form of the Liouville three-point function
(see the Appendix for the definition and a survey of functional properties
of the $\Upsilon _{b}$-function). The coefficient $c_{b}$ is a $b$-dependent
factor, independent of the quantum numbers $j_{\mu }$ and $\alpha _{\mu }$,
whose explicit value can be found in Ref. \cite{Tcross} and references
therein, though its explicit value is not important for our purpose here.
The relation between the indices $j_{\mu }$, that label the $SL(2,\mathbb{R}%
)_{k}$ representations, and the Liouville momenta $\alpha _{\mu }$ involves
a non-diagonal invertible transformation defined through%
\begin{equation}
2\alpha _{1}=b(j_{1}+j_{2}+j_{3}+j_{4}-1),\qquad 2\alpha
_{i}=b(j_{1}-j_{2}-j_{3}-j_{4}+2j_{i})+Q,\ i=\{2,3,4\}.  \label{FZ}
\end{equation}%
On the other hand, the relation between the WZNW level $k$ and the Liouville
parameter $b$ is given by 
\begin{equation*}
b^{-2}=k-2\in \mathbb{R}_{>0}.
\end{equation*}%
Now, let us make a remark on the KPZ scaling in (\ref{fia}): The parameter $%
\lambda $ is a free parameter of the WZNW theory and it can be regarded as
the mass of the two-dimensional black hole when the gauged $SL(2,\mathbb{R}%
)_{k}/U(1)$ is being considered. This is just a parameter and is free of
physical significance since it can be set to any positive value by means of
an appropriate rescaling of the zero-mode of the dilaton. Thus, we could
absorb the factor $\pi b^{2b^{2}}\Gamma (1-b^{2})/\Gamma (1+b^{2})$ in (\ref%
{fia2}) by simply shifting $\lambda $. This freedom can be used to simplify
the functional form of the correlators \cite{GKnotes}.

\subsection{The Stoyanovsky-Ribault-Teschner identity}

Analogously as to the case of the FZ map, the SRT map, discovered by
Stoyanovsky in \cite{S2000}, translates solutions of the KZ equation into
solutions of the BPZ equation. But this is, in some sense, more general.
Unlike the FZ map, the SRT map presents two advantages on its partner: the
first is that it involves a diagonal transformation between the quantum
numbers $j_{\mu }$ and $\alpha _{\mu }$; the second advantage, and more
important indeed, is that it holds for the case of the $N$-point KZ equation
for an arbitrary $N$. One of the original versions of the SRT map states the
correspondence between $N$-point functions in the WZNW theory and the $2N-2$%
-point function of the Liouville theory. Furthermore, this was generalized
by Ribault in order to connect any $N$-point functions in WZNW to $M$-point
functions in Liouville CFT with $N\leq M\leq 2N-2$. Let us describe such
generalized form below.

The Ribault formula reads 
\begin{equation*}
\langle \prod_{i=1}^{N}\Phi _{j_{i},m_{i},\bar{m}_{i}}^{\omega
_{i}}(z_{i})\rangle =\mathcal{N}_{k}(j_{1},...j_{N};m_{1},...m_{N})%
\prod_{r=1}^{M}\int d^{2}w_{r}\ \mathcal{F}%
_{k}(j_{1},...j_{N};m_{1},...m_{N}|z_{1},...z_{N};w_{1},...w_{M})\times
\end{equation*}%
\begin{equation}
\times \langle \prod_{t=1}^{N}V_{\alpha _{t}}(z_{t})\prod_{r=1}^{M}V_{-\frac{%
1}{2b}}(w_{r})\rangle \delta \left( \sum_{\mu =1}^{N}m_{\mu }-\frac{k}{2}%
M\right) \delta \left( \sum_{\mu =1}^{N}\overline{m}_{\mu }-\frac{k}{2}%
M\right) ,  \label{rt}
\end{equation}%
where the normalization factor is given by 
\begin{equation*}
\mathcal{N}_{k}(j_{1},...j_{N};m_{1},...m_{N})=\frac{2\pi ^{3-2N}b}{M!\
c_{k}^{M+2}}\prod_{i=1}^{N}\frac{c_{k}\ \Gamma (-m_{i}+j_{i})}{\Gamma
(1-j_{i}+\bar{m}_{i})},
\end{equation*}%
while the $z$-dependent function is 
\begin{equation*}
\mathcal{F}_{k}(j_{1},...j_{N};m_{1},...m_{N}|z_{1},...z_{N};w_{1},...w_{M})=%
\frac{\prod_{1\leq r<l}^{N}|z_{r}-z_{l}|^{k-2(m_{r}+m_{l}+\omega _{r}\omega
_{l}k/2+\omega _{l}m_{r}+\omega _{r}m_{l})}}{%
\prod_{1<r<l}^{M}|w_{r}-w_{l}|^{-k}\prod_{t=1}^{N}%
\prod_{r=1}^{M}|w_{r}-z_{t}|^{k-2m_{t}}}\times
\end{equation*}%
\begin{equation}
\times \frac{\prod_{1\leq r<l}^{N}(\bar{z}_{r}-\bar{z}_{l})^{m_{r}+m_{l}-%
\bar{m}_{r}-\bar{m}_{l}+\omega _{l}(m_{r}-\bar{m}_{r})+\omega _{r}(m_{l}-%
\bar{m}_{l})}}{\prod_{1<r<l}^{M}(\bar{w}_{r}-\bar{z}_{t})^{m_{t}-\bar{m}_{t}}%
}.  \label{F}
\end{equation}%
Here, the Liouville cosmological constant $\mu $ has been set to the value $%
\mu =b^{2}\pi ^{-2}$ for convenience; this is to make the KPZ scaling of
both sides of (\ref{rt}) match. It is also important to keep in mind the
presence of the free parameter $\lambda $ of the WZNW theory, and the fact
that this can still be set to an appropriate value in order to absorb the
powers of $\Gamma (b^{2})/\Gamma (1+b^{2})$ in the KPZ overall factor of
WZNW correlators \cite{GKnotes}. For our purpose, we do not need to focus
the attention on the specific value of the $j$-independent normalization
factor, which we wrote above just for completeness, being equal to $2\pi
^{3-2N}b/M!\ c_{k}^{M+2}$ and where $c_{k}$ represents a $k$-dependent
factor whose exact value is discussed in \cite{R2005} but, again, is not
relevant for us.

As in the case of the FZ map, the relation between the level $k$ of the WZNW
theory and the parameter $b$ of the Liouville theory is given by 
\begin{equation*}
b^{-2}=k-2\in \mathbb{R}_{>0},
\end{equation*}%
while the quantum numbers labeling the states of both theories are related
through 
\begin{equation}
\alpha _{i}=-bj_{i}+b+b^{-1}/2=b(k/2-j_{i})\ ,\ \ i=\{1,2,...N\}.
\label{simple}
\end{equation}%
We also have%
\begin{equation*}
s=-b^{-1}\sum_{i=1}^{N}\alpha _{i}+b^{-2}\frac{M}{2}+1+b^{-2}\ ,
\end{equation*}%
and, then, the total winding number is given by%
\begin{equation}
\sum_{i=1}^{N}\omega _{i}=M+2-N.  \label{ese}
\end{equation}%
This manifestly shows that scattering processes leading to the violation of
the total winding number conservation can occur in principle. \ In
particular, Ribault formula states that the four-point function involving
one flowed state of the sector $\omega =-1$ obeys%
\begin{equation*}
\langle \Phi _{J_{2},m_{2},\bar{m}_{2}}^{\omega _{2}=0}(0)\Phi _{J_{1},m_{1},%
\bar{m}_{1}}^{\omega _{1}=-1}(z)\Phi _{J_{3},m_{3},\bar{m}_{3}}^{\omega
_{3}=0}(1)\Phi _{J_{4},m_{4},\bar{m}_{4}}^{\omega _{4}=0}(\infty )\rangle =%
\widehat{c}_{b}\prod_{\mu =1}^{N}\frac{\ \Gamma (-m_{\mu }+J_{\mu })}{\Gamma
(1-J_{\mu }+\bar{m}_{\mu })}\times
\end{equation*}%
\begin{eqnarray*}
&&\times \delta (m_{1}+m_{2}+m_{3}-\frac{k}{2})\delta (\overline{m}_{1}+%
\overline{m}_{2}+\overline{m}_{3}-\frac{k}{2}%
)(z)^{k/2-m_{1}}(1-z)^{k/2-m_{1}}(\overline{z})^{k/2-\overline{m}_{1}}(1-%
\overline{z})^{k/2-\overline{m}_{1}}\times \\
&&\times \int d^{2}x(x)^{m_{2}-k/2}(\overline{x})^{\overline{m}%
_{2}-k/2}(1-x)^{m_{3}-k/2}(1-\overline{x})^{\overline{m}%
_{3}-k/2}(x-z)^{m_{1}-k/2}\ (\overline{x}-\overline{z})^{\overline{m}%
_{1}-k/2}\times
\end{eqnarray*}%
\begin{equation}
\times \langle V_{b\left( k/2-J_{2}\right) }(0)V_{b\left( k/2-J_{1}\right)
}(z)V_{b\left( k/2-J_{3}\right) }(1)V_{-\frac{1}{2b}}(x)V_{b\left(
k/2-J_{3}\right) }(\infty )\rangle ,  \label{iaf}
\end{equation}%
where $\widehat{c}_{b}$ is, again, a $J$-independent factor whose specific
value, $\widehat{c}_{b}=2\pi ^{-5}bc_{k}^{N-2-M}$, is known in terms of $%
c_{k}$, though not important for us. Notice that we have changed the
notation here, where we replaced $j_{\mu }$ by $J_{\mu }$; this will be
convenient later. Regarding the relation between $J_{\mu }$ and $\alpha
_{\mu }$, let us make some remarks here: Notice that, according to (\ref%
{simple}), the Liouville reflection $\alpha _{\mu }\rightarrow Q-\alpha
_{\mu }$ translates in terms of the WZNW correlators in doing the Weyl
reflection $j\rightarrow 1-j.$ This is a simple comment, but is also
important. In particular, because of (\ref{whatt}), this implies that the
correlation function involving the field $\Phi _{j}(x|z)$ and the one
involving the field $\Phi _{1-j}(x|z)$ are connected one to each other by
the overall factor $R_{b}(b(k/2-j))$. It is not difficult to see that this
simple affirmation leads directly to the exact expression for the WZNW
two-point function, for instance. This comment will become important below.

\section{The four-point function of winding strings}

\subsection{The idea}

The idea is to use Eq. (\ref{fia}) and Eq. (\ref{iaf}) in order to relate
the four-point function $\langle \Phi _{J_{1},m_{1},\bar{m}_{1}}^{\omega
_{1}=-1}(z_{1})$ $\prod_{i=2}^{4}\Phi _{J_{i},m_{i},\bar{m}_{i}}^{\omega
_{i}=0}(z_{i})\rangle $ with the four-point function $\langle \prod_{\mu
=1}^{4}\Phi _{j_{\mu }}(x_{\mu }|z_{\mu })\rangle $. The nexus will be the
Liouville five-point function $\langle \prod_{\mu =1}^{4}V_{\alpha _{\mu
}}(z_{\mu })V_{-1/2b}(x)\rangle $, appearing in the right hand side of both
expressions: this is connected to the first through the SRT map while is
connected to the second through the FZ map. After doing this, the second
step would be to \textquotedblleft diagonalize" the relation between indices 
$j_{\mu }$ and indices $J_{\mu }$. We will do this by using the preliminary
result of section 3.2. Notice that, according to Eq. (\ref{fia}) and Eq. (%
\ref{iaf}), the relation between momenta $J_{\mu }$ and $j_{\mu }$ is given
by (\ref{J}), below. The third step would be to perform a Fourier transform
in order to translate the operators $\Phi _{j_{\mu }}(x_{\mu }|z_{\mu })$ of
the $x$-basis into operators $\Phi _{j_{\mu },m_{\mu },\overline{m}_{\mu
}}(z_{\mu })$ of the $m$-basis.

\subsection{Non diagonal symmetry of the KZ equation}

As mentioned, the purpose here is to prove a functional relation obeyed by
two solutions of the KZ\ equation that will be useful further. Specifically,
we will see that four-point correlation functions with momenta $%
j_{1},...j_{4}$ have a simple expression in terms of the analogous quantity
with momenta $\widetilde{j}_{1},...\widetilde{j}_{4}$, being related through 
\begin{equation*}
2\widetilde{j}_{\mu }=\sum_{\nu =1}^{4}j_{\nu }-2j_{\mu }.
\end{equation*}%
To see this, let us begin by observing that, according to (\ref{FZ}), the
fact of performing the change $\widetilde{j}_{\mu }\rightarrow j_{\mu }$ for
all the indices $\mu =\{1,2,3,4\}$ translates in terms of the Liouville
momenta $\alpha _{\mu }$ in doing the change $\alpha _{i}\rightarrow
Q-\alpha _{i}$ for $i=\{2,3,4\},$ while keeping $\alpha _{1}$ unchanged.
Then, by taking into account the Liouville reflection symmetry (\ref{whatt})
and the dictionary (\ref{FZ}), we get\footnote{%
I thank Cecilia Garraffo for discussions on this formula.}%
\begin{equation*}
\left\langle \Phi _{j_{1}}(x|z)\Phi _{j_{2}}(0|0)\Phi _{j_{3}}(1|1)\Phi
_{j_{4}}(\infty |\infty )\right\rangle =\frac{%
F_{k}(j_{1},j_{2},j_{3},j_{4})X_{k}(j_{1},j_{2},j_{3},j_{4}|x,z)}{F_{k}(%
\widetilde{j}_{1},\widetilde{j}_{2},\widetilde{j}_{3},\widetilde{j}%
_{4})X_{k}(\widetilde{j}_{1},\widetilde{j}_{2},\widetilde{j}_{3},\widetilde{j%
}_{4}|x,z)}\prod_{i=2}^{4}R_{b}(\alpha _{i})\times
\end{equation*}%
\begin{equation}
\times \left\langle \Phi _{\widetilde{j}_{1}}(x|z)\Phi _{\widetilde{j}%
_{2}}(0|0)\Phi _{\widetilde{j}_{3}}(1|1)\Phi _{\widetilde{j}_{4}}(\infty
|\infty )\right\rangle  \label{wupa}
\end{equation}%
The factor $F_{k}(j_{1},j_{2},j_{3},j_{4})/F_{k}(\widetilde{j}_{1},%
\widetilde{j}_{2},\widetilde{j}_{3},\widetilde{j}_{4})$ involves a quotient
of products of $\Upsilon _{b}$-functions that, by making use of the formulae
in the Appendix (see (\ref{magicsuerte}) below), can be substantially
simplified once one takes into account the presence of the three factors $%
R_{b}(\alpha _{i})$ ($i=2,3,4$). Then, the results reads%
\begin{equation*}
\left\langle \Phi _{j_{1}}(x|z)\Phi _{j_{2}}(0|0)\Phi _{j_{3}}(1|1)\Phi
_{j_{4}}(\infty |\infty )\right\rangle =b^{2\mathcal{P}(j)}|x|^{2b^{-1}(Q-2%
\alpha _{2})}|1-x|^{2b^{-1}(Q-2\alpha
_{3})}|z|^{b^{2}(j_{1}+j_{2})^{2}-b^{2}(j_{3}+j_{4})^{2}}\times
\end{equation*}%
\begin{equation}
\times |1-z|^{b^{2}(j_{1}+j_{3})^{2}-b^{2}(j_{2}+j_{4})^{2}}\prod_{\mu
=1}^{4}\frac{\Upsilon _{b}(2j_{\mu }b-b)}{\Upsilon _{b}\left( b\sum_{\nu
=1}^{4}j_{\nu }-2j_{\mu }b-b\right) }\left\langle \Phi _{\widetilde{j}%
}(x|z)\Phi _{\widetilde{j}_{2}}(0|0)\Phi _{\widetilde{j}_{3}}(1|1)\Phi _{%
\widetilde{j}_{4}}(\infty |\infty )\right\rangle  \label{cardinal}
\end{equation}%
where $\mathcal{P}(j)$ is a polynomial in the indices $j_{\mu }$, namely $%
\mathcal{P}(j)=-3j_{1}+j_{2}+j_{3}+j_{4},$ though is not actually relevant
here. Since $j_{1}+j_{2}=\widetilde{j}_{3}+\widetilde{j}_{4}$ and $%
j_{1}+j_{3}=\widetilde{j}_{2}+\widetilde{j}_{4}$, one immediately notices
that the $\mathbb{Z}_{2}$-invariant form under the involution $j_{\mu
}\rightarrow \widetilde{j}_{\mu }$ is given by%
\begin{equation*}
\mathcal{I}_{k}^{\pm }(x,z)=Z_{k}^{\pm
}(j_{1},j_{2},j_{3},j_{4}|x,z)\left\langle \Phi _{j_{1}}(x|z)\Phi
_{j_{2}}(0|0)\Phi _{j_{3}}(1|1)\Phi _{j_{4}}(\infty |\infty )\right\rangle
\end{equation*}%
where%
\begin{eqnarray*}
Z_{k}^{+}(j_{1},j_{2},j_{3},j_{4}|x,z)
&=&b^{2(j_{1}-j_{2}-j_{3}-j_{4})}|x|^{j_{1}+j_{2}-j_{3}-j_{4}+k-1}|1-x|^{j_{1}-j_{2}+j_{3}-j_{4}+k-1}\times
\\
&&\times
|z|^{+b^{2}(j_{3}+j_{4})^{2}}|1-z|^{+b^{2}(j_{2}+j_{4})^{2}}\prod_{\mu
=1}^{4}\Upsilon _{b}^{-1}(2j_{\mu }b-b) \\
Z_{k}^{-}(j_{1},j_{2},j_{3},j_{4}|x,z)
&=&b^{2(j_{1}-j_{2}-j_{3}-j_{4})}|x|^{j_{1}+j_{2}-j_{3}-j_{4}+k-1}|1-x|^{j_{1}-j_{2}+j_{3}-j_{4}+k-1}\times
\\
&&\times
|z|^{-b^{2}(j_{1}+j_{2})^{2}}|1-z|^{-b^{2}(j_{1}+j_{3})^{2}}\prod_{\mu
=1}^{4}\Upsilon _{b}^{-1}(2j_{\mu }b-b).
\end{eqnarray*}%
Formula (\ref{cardinal}) will be useful in the next section. Similar
functional relations were studied in Ref. \cite{Yo2005,Andreev,Nichols}.
Now, let us move to the case of winding strings.

\subsection{The four-point function of winding strings}

In this subsection, as an application of our formula (\ref{cardinal}), we
will employ it to show that, by making use of both FZ and SRT maps, it is
feasible to write down a formula that expresses the winding violating
four-point functions in terms of the zero-winding four-point function. To do
this, let us begin by considering the FZ identity%
\begin{equation*}
\langle \Phi _{j_{2}}(0|0)\Phi _{1-j_{1}}(x|z)\Phi _{j_{3}}(1|1)\Phi
_{j_{4}}(\infty |\infty )\rangle
=F_{k}(1-j_{1},j_{2},j_{3},j_{4})X_{k}(1-j_{1},j_{2},j_{3},j_{4}|x,z)\times 
\end{equation*}%
\begin{equation}
\times \langle V_{\widehat{\alpha }_{2}}(0)V_{\widehat{\alpha }_{1}}(z)V_{%
\widehat{\alpha }_{3}}(1)V_{-\frac{1}{2b}}(x)V_{\widehat{\alpha }%
_{4}}(\infty )\rangle ,  \label{fiaty}
\end{equation}%
where the quantum numbers $\widehat{\alpha }_{\mu }$ and $j_{\mu }$ are then
related as it follows%
\begin{eqnarray*}
\widehat{\alpha }_{1} &=&\frac{b}{2}(-j_{1}+j_{2}+j_{3}+j_{4}),\qquad 
\widehat{\alpha }_{2}=\frac{b}{2}(-j_{1}+j_{2}-j_{3}-j_{4}+k), \\
\widehat{\alpha }_{3} &=&\frac{b}{2}(-j_{1}-j_{2}+j_{3}-j_{4}+k),\qquad 
\widehat{\alpha }_{4}=\frac{b}{2}(-j_{1}-j_{2}-j_{3}+j_{4}+k).
\end{eqnarray*}%
Now, let us also define quantum numbers $J_{\mu }$ as%
\begin{equation}
b(k/2-J_{\mu })=\widehat{\alpha }_{\mu }
\end{equation}%
for $\mu =\{1,2,3,4\}$. Then, by taking into account (\ref{whatt}), we have%
\begin{equation}
\langle V_{\widehat{\alpha }_{2}}(0)V_{\widehat{\alpha }_{1}}(z)V_{\widehat{%
\alpha }_{3}}(1)V_{-\frac{1}{2b}}(x)V_{\widehat{\alpha }_{4}}(\infty
)\rangle =\widetilde{A}_{k}(J_{1},J_{2},J_{3},J_{4}|x,z)\langle \Phi
_{j_{2}}(0|0)\Phi _{1-j_{1}}(x|z)\Phi _{j_{3}}(1|1)\Phi _{j_{4}}(\infty
|\infty )\rangle .  \label{fiatyu}
\end{equation}%
with%
\begin{equation}
\widetilde{A}%
_{k}(J_{1},J_{2},J_{3},J_{4}|x,z)=F_{k}^{-1}(1-j_{1},j_{2},j_{3},j_{4})X_{k}^{-1}(1-j_{1},j_{2},j_{3},j_{4}|x,z).
\label{volvio}
\end{equation}%
Consequently, we can write the following SRT identity%
\begin{equation*}
\langle \Phi _{J_{2},m_{2},\bar{m}_{2}}^{\omega _{2}=0}(0)\Phi _{J_{1},m_{1},%
\bar{m}_{1}}^{\omega _{1}=-1}(z)\Phi _{J_{3},m_{3},\bar{m}_{3}}^{\omega
_{3}=0}(1)\Phi _{J_{4},m_{4},\bar{m}_{4}}^{\omega _{4}=0}(\infty )\rangle =%
\widehat{c}_{b}\prod_{\mu =1}^{N}\frac{\ \Gamma (-m_{\mu }+J_{\mu })}{\Gamma
(1-J_{\mu }+\bar{m}_{\mu })}\times 
\end{equation*}%
\begin{eqnarray*}
&&\times \delta (m_{1}+m_{2}+m_{3}-k/2)\delta (\overline{m}_{1}+\overline{m}%
_{2}+\overline{m}_{3}-k/2)(z)^{k/2-m_{1}}(\overline{z})^{k/2-\overline{m}%
_{1}}(1-z)^{k/2-m_{1}}(1-\overline{z})^{k/2-\overline{m}_{1}} \\
&&\times \int d^{2}x(x)^{m_{2}-k/2}(1-x)^{m_{3}-k/2}(\overline{x})^{%
\overline{m}_{2}-k/2}(1-\overline{x})^{\overline{m}%
_{3}-k/2}(x-z)^{m_{1}-k/2}(\overline{x}-\overline{z})^{\overline{m}%
_{1}-k/2}\times 
\end{eqnarray*}%
\begin{equation}
\times \ \langle V_{\widehat{\alpha }_{2}}(0)V_{\widehat{\alpha }_{1}}(z)V_{%
\widehat{\alpha }_{3}}(1)V_{-\frac{1}{2b}}(x)V_{\widehat{\alpha }%
_{4}}(\infty )\rangle .  \label{iafyy}
\end{equation}%
Hence, plugging (\ref{fiatyu}) into (\ref{iafyy}), we find%
\begin{equation*}
\left\langle \Phi _{J_{1},m_{1},,\overline{m}_{1}}^{\omega _{1}=-1}(z)\Phi
_{J_{2},m_{2},,\overline{m}_{2}}^{\omega _{1}=0}(0)\Phi _{J_{3},m_{3},%
\overline{m}_{3}}^{\omega _{1}=0}(1)\Phi _{J_{4},m_{4},\overline{m}%
_{4}}^{\omega _{4}=0}(\infty )\right\rangle =\prod_{\mu =1}^{4}\frac{\Gamma
(J_{\mu }-m_{\mu })}{\Gamma (1-J_{\mu }+\overline{m}_{\mu })}\times 
\end{equation*}%
\begin{eqnarray}
&&\times \delta (m_{1}+m_{2}+m_{3}-k/2)\delta (\overline{m}_{1}+\overline{m}%
_{2}+\overline{m}_{3}-k/2)(z)^{-m_{1}}(\overline{z})^{-\overline{m}%
_{1}}(1-z)^{-m_{1}}(1-\overline{z})^{-\overline{m}_{1}}\times   \notag \\
&&\times \int d^{2}x\ (x-z)^{m_{1}}(\overline{x}-\overline{z})^{\overline{m}%
_{1}}(x)^{m_{2}}(\overline{x})^{\overline{m}_{2}}(1-x)^{m_{3}}(1-\overline{x}%
)^{\overline{m}_{3}}\ A_{k}(J_{1},J_{2},J_{3},J_{4}|x,z)\times   \notag \\
&&\times \left\langle \Phi _{1-j_{1}}(x|z)\Phi _{j_{2}}(0|0)\Phi
_{j_{3}}(1|1)\Phi _{j_{4}}(\infty |\infty )\right\rangle ,  \label{A}
\end{eqnarray}%
where we simply defined%
\begin{equation*}
A_{k}(J_{1},J_{2},J_{3},J_{4}|x,z)=\left| \frac{z\ (1-z)}{x\ (1-x)(z-x)}%
\right| ^{k}\widetilde{A}_{k}(J_{1},J_{2},J_{3},J_{4}|x,z)
\end{equation*}%
and where, according to the definitions above, the indices $J_{\mu }$ and $%
j_{\mu }$ turn out to be related by%
\begin{equation}
2J_{1}=k+j_{1}-j_{2}-j_{3}-j_{4},\qquad 2J_{i}=j_{1}+j_{2}+j_{3}+j_{4}-2j_{i}
\label{J}
\end{equation}%
for $i=\{2,3,4\}$. Then, it is feasible to show that%
\begin{equation*}
A_{k}(J_{1},J_{2},J_{3},J_{4}|x,z)\sim \frac{\Upsilon _{b}(2b(k/2-J_{1}))}{%
\Upsilon _{b}(b(k/2+1-J_{1}-J_{2}-J_{3}-J_{4}))}%
|x|^{-2J_{2}}|1-x|^{-2J_{3}}|x-z|^{-2J_{1}}\times 
\end{equation*}%
\ \qquad \qquad 
\begin{equation}
\times \prod_{i=2}^{4}\frac{\Upsilon _{b}(b(2J_{i}-1))}{\Upsilon
_{b}(b(2J_{i}-J_{1}-J_{2}-J_{3}-J_{4}+k/2-1))}|z|^{k+4b^{2}(1-j_{1})j_{2}-4%
\widehat{\alpha }_{1}\widehat{\alpha }_{2}}|1-z|^{k+4b^{2}(1-j_{1})j_{3}-4%
\widehat{\alpha }_{1}\widehat{\alpha }_{3}}.  \label{elA}
\end{equation}%
where the symbol $\sim $ stands just because we are not writing the $k$%
-dependent overall factor and the precise KPZ\ scaling here (this can be
directly read from Eq. (\ref{volvio}) if necessary). For short, in the last
equation we have employed the notation%
\begin{equation}
2\widehat{\alpha }_{1}=b(k/2-J_{1})=b(-j_{1}+j_{2}+j_{3}+j_{4}),\qquad 2%
\widehat{\alpha }_{i}=b(k/2-J_{i})=b(k+2j_{i}-j_{1}-j_{2}-j_{3}-j_{4}),
\label{alfahat}
\end{equation}%
and one could also find convenient to define%
\begin{equation}
2\alpha _{1}=b(j_{1}+j_{2}+j_{3}+j_{4}-1),\qquad 2\alpha
_{i}=b(j_{1}-j_{2}-j_{3}-j_{4}+2j_{i}+k-1).  \label{alfa}
\end{equation}%
Identity (\ref{A}) does already express the winding violating correlators in
terms of the conservative ones. However, such expression is still not fully
satisfactory since the right hand side of (\ref{A}) contains a four-point
correlator that, instead of involving states with momenta $J_{\mu }$,
involves states with momenta $j_{\mu }$, $\mu =\{1,2,3,4\}$. Here is where
our result of subsection 3.2. becomes useful, since the following step would
be to relate the correlator in the $j_{\mu }$-basis with the one that is
diagonal in the $J_{\mu }$-basis. This can be simply achieved by making use
of the equation (\ref{cardinal}) and by taking into account (\ref{J}),
yielding 
\begin{equation*}
\left\langle \Phi _{j_{1}}(x|z)\Phi _{j_{2}}(0|0)\Phi _{j_{3}}(1|1)\Phi
_{j_{4}}(\infty |\infty )\right\rangle
=B_{k}(J_{1},J_{2},J_{3},J_{4}|x,z)\times 
\end{equation*}%
\begin{equation}
\times \left\langle \Phi _{\frac{k}{2}-J_{1}}(x|z)\Phi _{J_{2}}(0|0)\Phi
_{J_{3}}(1|1)\Phi _{J_{4}}(\infty |\infty )\right\rangle ,  \label{B}
\end{equation}%
with the normalization factor being%
\begin{equation*}
B_{k}(J_{1},J_{2},J_{3},J_{4}|x,z)=b^{2\widehat{\mathcal{P}}%
(j)}|z|^{b^{2}(J_{3}+J_{4})^{2}-b^{2}(k/2-J_{1}+J_{2})^{2}}|1-z|^{b^{2}(J_{2}+J_{4})^{2}-b^{2}(k/2-J_{1}+J_{3})^{2}}\times 
\end{equation*}%
\begin{eqnarray}
&&\times \frac{\Upsilon _{b}(b(J_{1}+J_{2}+J_{3}+J_{4}-k/2-1))}{\Upsilon
_{b}(b(k-2J_{1}-1))}\prod_{i=2}^{4}\frac{\Upsilon
_{b}(b(2J_{i}-J_{1}-J_{2}-J_{3}-J_{4}+k/2-1))}{\Upsilon _{b}(b(2J_{i}-1))}%
\times   \notag \\
&&\times
|x|^{k-2(J_{1}-J_{2}+J_{3}+J_{4})}|1-x|^{k-2(J_{1}+J_{2}-J_{3}+J_{4})};
\label{elB}
\end{eqnarray}%
here, $\widehat{\mathcal{P}}(j)$ is again a polynomial in the indices $%
j_{\mu }$ that is related to the $\mathcal{P}(j)$ in (\ref{cardinal})
through the replacement $j_{i}\rightarrow J_{i}$, $i=\{2,3,4\}$, $%
j_{1}\rightarrow k/2-J_{1}$. In writing down (\ref{elB}) we used the fact
that, as in the previous subsection, the combination of the $\Upsilon _{b}$%
-functions and the $\Gamma $-functions in $R_{b}(\alpha _{i})$ leads to a
very simple expression. In this case, it comes from 
\begin{equation*}
\frac{\Upsilon _{b}(b(j_{1}-j_{2}-j_{3}-j_{4}+2j_{i}))}{\Upsilon
_{b}(b(-j_{1}+j_{2}+j_{3}+j_{4}-2j_{i}))R_{b}\left( \frac{Q}{2}-\frac{b}{2}%
(j_{1}-j_{2}-j_{3}-j_{4}+2j_{i})\right) }=
\end{equation*}%
\begin{equation}
=\left( \pi b^{2(b^{2}-1)}\frac{\Gamma (1-b^{2})}{\Gamma (1+b^{2})}\right)
^{J_{1}-J_{2}-J_{3}-J_{4}+2J_{i}-k/2}.  \label{magicsuerte}
\end{equation}%
Hence, up to the Weyl reflection $j_{1}\rightarrow 1-j_{1}$ that we discuss
below, equations (\ref{B}) and (\ref{A}) represent the result we wanted to
prove: the four-point function involving one winding state of the sector $%
\omega =-1$ admits to be expressed in terms of the four-point function of
non-winding states. Let us discuss the final form of this result in the
following paragraphs (see (\ref{DaleR}) below), where we address the
comparison with the case of the three-point function. We will also comment
on possible applications and conclude with some remarks in the following
section.

\subsection{Analogy with the case of the three-point function}

The three-point function that includes one spectral flowed state of the
sector $\omega =-1$ can be written in terms of the structure constant of
three non-flowed states \cite{FZZ, MO3}. In the subsection above we proved a
similar relation at the level of the four-point functions. Then, the natural
question we want to address now is whether both relations are connected in
some way. As we will see below, these are indeed analogous. Actually,
expressions (\ref{A}) and (\ref{B}), once considered together, correspond to
the generalization of the formula that holds for the three-point structure
constants. First, one of the aspects one notices in the formulae above is
that the first operator in the right hand side of (\ref{B}) represents the
state of momentum $\frac{k}{2}-J_{1}$, instead that of momentum $J_{1}$.
Actually, this should not be a surprise since it precisely resembles what
happens at the level of the three-point function, where the violating
winding correlator 
\begin{equation*}
\left\langle \Phi _{J_{1},m_{1},,\overline{m}_{1}}^{\omega _{1}=-1}(0)\Phi
_{J_{2},m_{2},,\overline{m}_{2}}^{\omega _{2}=0}(1)\Phi _{J_{3},m_{3},%
\overline{m}_{3}}^{\omega _{3}=0}(\infty )\right\rangle
\end{equation*}%
turns out to be proportional to the integral of the conservative structure
constant that corresponds to%
\begin{equation*}
\sim B^{-1}(k/2-J_{1})\left\langle \Phi _{k/2-J_{1}}(0|0)\Phi
_{J_{2}}(1|1)\Phi _{J_{3}}(\infty |\infty )\right\rangle .
\end{equation*}%
The same association between momenta $J_{1}$ and $k/2-J_{1}$ occurs for the
four-point functions here. Actually, such relation between states with
quantum numbers $\omega =0$, $J$ and those with $\omega =\pm 1$, $k/2-J$ is
understood from the algebraic point of view: This concerns the
identification between the discrete series of the $SL(2,\mathbb{R})_{k}$
representations by means of the spectral flow automorphism. This is related
to the fact hat both states with $\omega =0$ and $|\omega |=1$ satisfy
similar KZ\ equations. Besides, this is related to the fact that the OPE
between the vertex operator $\Phi _{J}(z)$ and the spectral flow operator $%
\Phi _{k/2}(w)$, which is necessary to provide the winding $\omega _{1}=1$
to the first, yields the single string contribution $\Phi _{k/2-J}(z)$. The
involution $J\rightarrow k/2-J$, as a symmetry of the KZ equation, was
studied in the first part of our work, \cite{Yo2005}. As it was detailed in
Ref. \cite{MO3}, when presenting the original derivation of \cite{FZZ}, the
proportionality factor connecting both violating winding and non-violating
winding three-point correlators is basically given by the reflection
coefficient $B(J_{1})\sim B^{-1}(k/2-J_{1})$ of the $SL(2,\mathbb{R})_{k}$
WZNW model (see the formula above). This quantity has the form%
\begin{equation}
B(J)=\frac{1}{\pi b^{2}}\left( \lambda \pi \frac{\Gamma \left(
1-b^{2}\right) }{\Gamma \left( 1+b^{2}\right) }\right) ^{1-2J}\frac{\Gamma
\left( 1+b^{2}-2Jb^{2}\right) }{\Gamma \left( 2Jb^{2}-b^{2}\right) },
\label{bB}
\end{equation}%
and therefor%
\begin{equation*}
B(J)=\frac{1}{\pi ^{2}b^{4}}\left( \frac{1}{\lambda \pi }\frac{\Gamma \left(
1+b^{2}\right) }{\Gamma \left( 1-b^{2}\right) }\right)
^{b^{-2}}B^{-1}(k/2-J).
\end{equation*}%
This permits to show that the picture for the three-point functions turns
out to be similar to the one we are obtaining here for the four-point
functions. Actually, this is not only manifested in the shifting $%
J_{1}\rightarrow k/2-J_{1}$, but also in the fact that, when combining both (%
\ref{A}) and (\ref{B}), a factor $\frac{\Upsilon _{b}(2b(k/2-J_{1}))}{%
\Upsilon _{b}(2b(k/2-J_{1})-b)}$ also arises in the product between $%
A_{k}(J_{1},J_{2},J_{3},J_{4}|x,z)$ and $B_{k}(J_{1},J_{2},J_{3},J_{4}|x,z)$%
. This factor is precisely proportional to \thinspace $(2J-1)B(J)$; thus,
this is actually in equal footing as to how the WZNW reflection factor
stands in the relation between violating and non-violating three-point
correlators.

On the other hand, some nice cancellations occur when combining expressions (%
\ref{A}) and (\ref{B}). First, we observed that, instead of the operator $%
\Phi _{j_{1}}(x|z)$, the correlator in the right hand side of (\ref{A})
involves its Weyl reflected operator $\Phi _{1-j_{1}}(x|z)$; and, as we
commented at the end of section 2, this implies that a factor $%
R_{b}^{-1}(b(k/2-j_{1}))$ has to be included as well in order to plug (\ref%
{A}) into (\ref{B}). Then, such reflection coefficient is eventually
simplified due to another contribution standing when multiplying $%
A_{k}(J_{1},J_{2},J_{3},J_{4}|x,z)$ and $B_{k}(J_{1},J_{2},J_{3},J_{4}|x,z)$%
; namely, 
\begin{equation*}
\frac{\Upsilon _{b}(b(J_{1}+J_{2}+J_{3}+J_{4}-k/2-1))}{\Upsilon
_{b}(-b(J_{1}+J_{2}+J_{3}+J_{4}-k/2-1))}=\frac{\Upsilon _{b}(2j_{1}b-b)}{%
\Upsilon _{b}(b-2j_{1}b)}\sim R_{b}(b(k/2-j_{1})).
\end{equation*}%
Moreover, these are not all the cancellations that take place. Let us also
observe that three factors of the form $\frac{\Upsilon _{b}(b(2J_{i}-1))}{%
\Upsilon _{b}(b(2J_{i}-J_{1}-J_{2}-J_{3}-J_{4}+k/2-1))}$ (with $i=\{2,3,4\}$%
) and their respective inverses are mutually cancelled as well. These come
from the second line of Eq. (\ref{elA}) and the second line of Eq. (\ref{elB}%
), respectively. Finally, equations (\ref{A}) and (\ref{B}), considered
together, leads to the following expression%
\begin{equation*}
\left\langle \Phi _{J_{1},m_{1},,\overline{m}_{1}}^{\omega _{1}=-1}(z)\Phi
_{J_{2},m_{2},,\overline{m}_{2}}^{\omega _{1}=0}(0)\Phi _{J_{3},m_{3},%
\overline{m}_{3}}^{\omega _{1}=0}(1)\Phi _{J_{4},m_{4},\overline{m}%
_{4}}^{\omega _{4}=0}(\infty )\right\rangle \sim B(J_{1})\prod_{\mu =1}^{4}%
\frac{\Gamma (J_{\mu }-m_{\mu })}{\Gamma (1-J_{\mu }+\overline{m}_{\mu })}%
\times 
\end{equation*}%
\begin{eqnarray}
&&\int d^{2}x\ (x-z)^{m_{1}+\Delta _{1}}(\overline{x}-\overline{z})^{%
\overline{m}_{1}+\Delta _{1}}\prod_{i=1}^{2}(z-z_{i})^{-m_{i}+\Delta _{i}}(%
\overline{z}-\overline{z_{i}})^{-\overline{m}_{i}+\Delta
_{i}}(x-x_{i})^{m_{i}+\widetilde{\Delta }_{i}}(\overline{x}-\overline{x}%
_{i})^{\overline{m}_{i}+\widetilde{\Delta }_{i}}\times   \notag \\
&&\times \left\langle \Phi _{\frac{k}{2}-J_{1}}(x|z)\Phi _{J_{2}}(0|0)\Phi
_{J_{3}}(1|1)\Phi _{J_{4}}(\infty |\infty )\right\rangle \delta
(m_{1}+m_{2}+m_{3}-k/2)\delta (\overline{m}_{1}+\overline{m}_{2}+\overline{m}%
_{3}-k/2),  \notag \\
&&  \label{DaleR}
\end{eqnarray}%
where $B(J_{1})$ is the WZNW reflection coefficient (\ref{bB}), while the
symbol $\sim $ stands for some $k$-dependent overall factor. For short, in
this expression we denoted $x_{2}=z_{2}=0$ and $x_{3}=z_{3}=1$, and
exponents $\Delta _{1,2,3}$ and $\widetilde{\Delta }_{2,3}$ refer to $J$%
-dependent ($m$-independent) linear combinations that are directly given by
the exponents arising in (\ref{A}) and (\ref{B}). Expression (\ref{DaleR})
represents the main result here, and it turns out to be actually analogous
to the formula connecting winding violating three-point functions to those
involving merely three non-winding states. Namely, we showed that the $%
SL(2,R)_{k}$ WZNW four-point function%
\begin{equation*}
\left\langle \Phi _{J_{1},m_{1},,\overline{m}_{1}}^{\omega _{1}=-1}(z)\Phi
_{J_{2},m_{2},,\overline{m}_{2}}^{\omega _{1}=0}(0)\Phi _{J_{3},m_{3},%
\overline{m}_{3}}^{\omega _{1}=0}(1)\Phi _{J_{4},m_{4},\overline{m}%
_{4}}^{\omega _{4}=0}(\infty )\right\rangle ,
\end{equation*}%
while involving one state of the spectral flowed sector $\omega _{1}=-1,$
can be expressed in terms of the integral of the four-point function%
\begin{equation*}
\sim B^{-1}(k/2-J_{1})\left\langle \Phi _{\frac{k}{2}-J_{1}}(x|z)\Phi
_{J_{2}}(0|0)\Phi _{J_{3}}(1|1)\Phi _{J_{4}}(\infty |\infty )\right\rangle ,
\end{equation*}%
defined in terms of merely non-spectral flowed states $\omega _{\mu }=0$, $%
\mu =\{1,2,3,4\}$. The integration over the complex variable $x$ clearly
stands for the requirement of the Fourier transform when changing to the $m$%
-basis. Furthermore, it is worth pointing out that the factor%
\begin{equation*}
\prod_{\mu =1}^{4}\frac{\Gamma (J_{\mu }-m_{\mu })}{\Gamma (1-J_{\mu }+%
\overline{m}_{\mu })}
\end{equation*}%
is also present here, completing the analogy with the three-point function
case. The fact that a similar pattern is found at the level of the
four-point correlators opens a window that would permit to gain information
about the violating winding four-point functions by making use of all what
is known about the WZNW conservative amplitudes.

\subsection{Further applications}

So far, we presented a concise application of our results of \cite{Yo2005},
by showing that the symmetries of the KZ equation, that were inferred by
means of its relation to the BPZ equation in Liouville theory, lead to a
relation between violating and conserving winding amplitudes in $AdS_{3}$
string theory. Besides, there exist some further application of the formula (%
\ref{A}) we obtained here: One of these feasible applications concerns the
integral representation of the four-point conformal blocks and the
factorization \textit{ansatz}. Actually, let us return to (\ref{antonia})
and expand the chiral conformal blocks\ as it follows%
\begin{equation}
{\mathcal{G}}%
_{j_{1},j_{2,}j,j_{3},j_{4}}(x|z)=z^{h_{j}-h_{j_{1}}-h_{j_{2}}}x^{j-j_{1}-j_{2}}\sum_{n=0}^{\infty }%
{\mathcal{G}}_{j_{1},j_{2,}j,j_{3},j_{4}}^{(n)}(x)z^{n},  \label{niato}
\end{equation}%
where, now, $j$ acts as an internal index that labels different solutions to
the differential equation. In the \textit{stringy} interpretation it is
feasible to assign physical meaning to such index: this is the one
parameterizing the intermediate states interchanged in a given four-point
scattering process. Then, the integration over the internal index $j$ stands
in order to include all the contributions of the conformal blocks (\ref%
{antonia}) to the four-point amplitude (\ref{recorda}). By replacing (\ref%
{niato}) into the KZ equation one finds that the leading term contribution
in the $z$-power expansion obeys the hypergeometric differential equation.
For instance, this permits to analyze the monodromy properties at $x=z$ for
leading orders in the large $1/x$ limit. Studying this regime turns out to
be important to fully understand the analytic structure of the four-point
function \cite{MO3}. For such purpose, it is useful to study the solution in
the vicinity of the point $x=z$. By assuming the extension of the
expressions above to the $SL(2,\mathbb{R})$ case, and by deforming the
contour of integration as ${\mathcal{C}}\rightarrow \frac{k}{2}-{\mathcal{C}}%
=\frac{k-1}{2}-i\mathbb{R}$, one can see that, once the integral over $j$ is
performed, the solution is actually monodromy invariant at $z=x$. Then, the
leading contribution of the solution can be written as it follows \cite{MO3}%
\begin{equation*}
|f_{j_{1},j_{2},j_{3},j_{4}}(x,z)|^{2}=\frac{1}{2}\int_{\frac{k-1}{2}+i%
\mathbb{R}%
}dj|x|^{2(h_{j}-h_{j_{1}}-h_{j_{2}}-j+j_{1}+j_{2})}|zx^{-1}|^{2(h_{j}-h_{j_{1}}-h_{j_{2}})}%
\frac{C(j_{1},j_{2},j)C(j,j_{3},j_{4})}{B(j)}\times
\end{equation*}%
\begin{equation*}
\times |F(j_{1}+j_{2}-j,j_{3}+j_{4}-j,k-2j,zx^{-1})|^{2}\left( 1+{\mathcal{O}%
}\left( z^{-1}x\right) \right) +2\pi i\sum_{\{x_{i}\}}Res_{(x=x_{i})}
\end{equation*}%
where $F(a,b,c;d)$ is the hypergeometric function, and where $\{x_{i}\}$
refers to the set of poles located in the region $1<Re(2j)<k-1$. These poles
take the form $j-j_{1}-j_{2}\in \mathbb{N}$ if the constraint $%
\sum_{i=1}^{4}j_{i}<k$ is assumed (see \cite{MO3} for the details of the
construction and the issue of the integration over the complex variables $z$
and $x$).\ Taking all this into account, one could try to plug the
expression for the conformal blocks (\ref{antonia}) in the equation we wrote
in (\ref{A}). This would lead to an integral realization of the winding
violating four-point function. Moreover, such a realization could be then
evaluated in a particular case $J_{4}=k/2$ and, by means of the prescription
of \cite{FZZ}, be used to compute the three-point function $\langle \Phi
_{J_{1},m_{1},\bar{m}_{1}}^{\omega _{1}=-1}\Phi _{J_{2},m_{2},\bar{m}%
_{2}}^{\omega _{2}=0}\Phi _{J_{3},m_{3},\bar{m}_{3}}^{\omega _{3}=+1}\rangle
,$ with two winding states of the sectors $\omega =+1$ and $\omega =-1$.
This idea does deserve to be explored in future work\footnote{%
I thank Pablo Minces for pointing me out this possible application.}.

Other possible analysis that can be done regards the Coulomb gas-like
integral representation recently presented in \cite{Yo2005b,Yo2006}. There,
it was proven that the SRT map can be thought of as a free field
representation of the $SL(2,\mathbb{R})_{k}$ model in terms of the action%
\begin{equation}
S=\frac{1}{4\pi }\int d^{2}z\left( -\partial \varphi \bar{\partial}\varphi
+QR\varphi +\mu e^{\sqrt{2}b\varphi }\right) +S_{M},
\end{equation}%
where the specific model representing the \textquotedblleft matter
sector\textquotedblright\ $S_{M}$\ corresponds to a $c<1$ conformal field
theory defined by the action%
\begin{equation*}
S_{M}=\frac{1}{4\pi }\int d^{2}z\left( \partial X^{0}\bar{\partial}%
X^{0}-\partial X^{1}\bar{\partial}X^{1}-i\sqrt{k}RX^{1}\right) .
\end{equation*}%
This free field representation leads to an integral expression for the
four-point function when the winding number conservation is being maximally
violated. On the other hand, for the cases where the winding is conserved,
the action $S_{M}$ has to be perturbed by introducing a new term of the form%
\begin{equation}
\mathcal{O}=\int d^{2}z\ e^{-\sqrt{\frac{k-2}{2}}\varphi (z)+i\sqrt{\frac{k}{%
2}}X^{1}(z)}.
\end{equation}%
This is a perturbation, represented by a primary operator of the matter
sector and properly dressed with the coupling to the Liouville field in
order to turn it into a marginal deformation. It could be interesting to
check the formulas (\ref{A})-(\ref{B}) in terms of this representation.

\section{Concluding remarks}

As an application of our \cite{Yo2005}, in this brief note we proved a new
relation existing between the four-point function of winding strings in $%
AdS_{3}$ (spectral flowed states of the $SL(2,\mathbb{R})_{k}$ WZNW model)
and the four-point function of non-winding strings (non-flowed $SL(2,\mathbb{%
R})_{k}$ states). We showed that the former admits a simple expression in
terms of the last, which was studied in more detail in the literature. The
fact that such a simple expression exits\ is mainly due to two facts: first,
the non-diagonal $\mathbb{Z}_{2}$ symmetry of the KZ equation realized by (%
\ref{cardinal}); secondly, to the fact that correlators of the WZNW model
are connected to those of the Liouville theory in more than one way.

To be more precise, here we have shown the relation that connects the
four-point WZNW correlation function involving one $\omega =-1$ spectral
flowed state to the four-point function of non-spectral flowed states. This
consequently relates the scattering amplitude involving one winding string
state in $AdS_{3}$ with the analogous observable for non-winding strings.
Such relation is realized by Eq. (\ref{DaleR}), and is reminiscent of the
relation obeyed by the $SL(2,\mathbb{R})_{k}$ structure constants. On the
other hand, it is likely that the four-point function involving one state in
the sector $\omega =-1$ corresponds to a limiting procedure of a WZNW
five-point function involving one spectral flow operator of momentum $%
J_{5}=k/2$. Consequently, it is plausible that the connection manifested by
equations (\ref{A}) and (\ref{B}) could then be obtained in an alternative
way; for instance, by directly studying the operator product expansion of
operators $\Phi _{J_{1}}(x_{1}|z_{1})\Phi _{k/2}(x_{2}|z_{2})$ in the
coincidence limit in the five-point conformal blocks. In such case, our
result would be seen from a different perspective since, as far as our
derivation of (\ref{A}) and (\ref{B}) is invertible, this would lead to the
possibility of deriving the FZ map from a particular case of the SRT map.
This was, indeed, one of the main motivations we had for studying this,
though the operator product expansion involving an additional spectral flow
operator (a fifth operator) turns out to be complicated enough and thus we
leave it for the future. Conversely, as far as the relation between both
maps is not trivial, as it is emphasized in \cite{RT2005}, our way of
proving the connection between violating winding correlators and conserving
winding correlators turns out to be an ingenious trick. The conciseness of
such a deduction becomes particularly evident once one gets familiarized
with the complicated definition of the action of spectral flow operator in
the $x$-basis. Moreover, even in the $m$-basis the computation turns out to
be simplified by our method because no explicit reference to the decoupling
equation of the degenerate state $J=k/2$ was needed at all, since it is was
already encoded in the Ribault formula (\ref{iaf}). Our hope is that the
result of this brief paper might be useful in working out the details of the 
$SL(2,\mathbb{R})_{k}$ WZNW four-point functions, which is our main
challenge within this line of research.

\begin{equation*}
\end{equation*}%
\textbf{Acknowledgement:} First, I would like to thank Cecilia Garraffo for
collaboration in the computations of the section 3. I am also grateful to
Universit\'{e} Libre de Bruxelles, and specially thank Glenn Barnich, Frank
Ferrari, Marc Henneaux and Mauricio Leston for their hospitality during my
stay, where the first part of this work was done. I also thank Pablo Minces
for his interest on this work and for very interesting suggestions about
possible applications. Last, I thank the Physics Department of New York
University for the hospitality during my stay, where the revised version of
the papers was finished. This work was partially supported by Universidad de
Buenos Aires and CONICET, Argentina.

\section*{Appendix: The function $\Upsilon _{b}(x)$}

The function $\Upsilon _{b}(x)$ was introduced by Zamolodchikov and
Zamoldchikov in Ref. \cite{ZZ}, and is defined as it follows%
\begin{equation}
\log \Upsilon _{b}(x)=\frac{1}{4}\int_{\mathbb{R}_{>0}}\frac{d\tau }{\tau }%
\left( (Q-2x)^{2}e^{-\tau }-\frac{\sinh ^{2}(\frac{\tau }{4}(Q-2x))}{\sinh (%
\frac{b\tau }{2})\sinh (\frac{\tau }{2b})}\right) ,  \label{upsilondefintion}
\end{equation}%
where $Q=b+b^{-1}$, being $b\in \mathbb{R}_{>0}.$ It\ also admits a
definition in terms of a limiting procedure involving the double Barnes $%
\Gamma _{2}$-function, though we did not find such a relation necessary
here. This function has its zeros at the points 
\begin{eqnarray*}
x &=&mb+nb^{-1} \\
x &=&-(m+1)b-(n+1)b^{-1}
\end{eqnarray*}%
for any pair of positive integers $m,n\in \mathbb{Z}_{>0}$. From (\ref%
{upsilondefintion}), it turns out to be evident that this function is
symmetric under the inversion of the parameter $b\rightarrow 1/b$, namely%
\begin{equation}
\Upsilon _{b}(x)=\Upsilon _{1/b}(x).
\end{equation}%
This is the first of a list of nice functional properties of this function.
The second identity we find useful is the reflection property%
\begin{equation}
\Upsilon _{b}(x)=\Upsilon _{b}(Q-x),  \label{Ur}
\end{equation}%
which is keeps track of the reflection symmetry of Liouville structure
constants when these are written in terms of $\Upsilon _{b}(x)$. Besides, (%
\ref{upsilondefintion}) also presents the following properties under fixed
translations%
\begin{equation}
\Upsilon _{b}(x+b^{\pm 1})=\Upsilon _{b}(x)\frac{\Gamma (b^{\pm 1}x)}{\Gamma
(1-b^{\pm 1}x)}b^{\pm 1\mp 2b^{\pm 1}x}.
\end{equation}%
The above identities are then gathered in the equation%
\begin{equation}
\Upsilon _{b}(Q\mp x)=\pm \Upsilon _{b}(x)\frac{\Gamma (bx)\Gamma (b^{-1}x)}{%
\Gamma (\pm bx)\Gamma (\pm b^{-1}x)}b^{2x(b^{\pm 1}-b)},
\end{equation}%
which, in particular, includes (\ref{Ur}). The relations above also permits
to prove that the following relation holds%
\begin{equation}
\Upsilon _{b}(x)=\Upsilon _{b}(-x)b^{2x(b-b^{-1})}\frac{\Gamma (-bx)\Gamma
(-b^{-1}x)}{\Gamma (bx)\Gamma (b^{-1}x)}.
\end{equation}%
All these relations were employed through our computation.

\end{document}